\documentclass[12pt]{article}
\usepackage{amsmath,amssymb,bbm}
\begin{document}

\begin{center}

\vskip 33.0cm

{\bf \large {A comment on $p$-branes of ($p+3$)$\rm{d}$ string theory}}

\vskip 1.5cm

{Eun Kyung Park\footnote{E-mail : ekpark@ks.ac.kr} and Pyung Seong
Kwon\footnote{E-mail : bskwon@ks.ac.kr}

Department of Physics, Kyungsung University, Pusan 608-736, Korea}

\end{center}
\thispagestyle{empty}

\vskip 2.0cm
\begin{abstract}
 We argue that in ($p+3$)d string theory the existence of NS-NS type $p$-brane with negative tension is essential to obtain background geometry $R_2$ or $R_2 / Z_n$ on the transverse dimensions, and the usual codimension-2 brane solutions with these background geometries already contain the negative tension NS-brane implicity in their ansatz. Such an argument leads us, in the context of brane world scenarios, to a conjecture that true background $p$-brane immanent in our spacetime may perhaps be NS-NS type brane, rather than D-brane.
\end{abstract}

\vskip 9.0cm
\medskip
\begin{center}
\end{center}

\newpage
\setcounter{page}{1}
\setcounter{footnote}{0}

\baselineskip 6.0mm

One of the most exciting objects of the string theory may be the
brane, a topologically stable extended object with its own charge.
There are two kinds of branes in string theory. The first one is
D-brane \cite{1,2}, which is known to carry R-R charge. The
D-brane is particularly interesting because it is conjectured that
our universe may be a stack of D-branes with standard
model(SM) fields living on it \cite{1,3}. The next one is NS-brane
\cite{2,4}, which is known to carry NS-NS charge. The NS-brane is
also interesting because it sometimes acts as a background
brane \cite{5,6} on which the D-branes (SM-brane) are to be set, or
in some cases it has its own SM-spectrum \cite{7}. In particular,
it plays an important role in the theories like "Little String
Theory" \cite{6,7}.

As mentioned above, D-branes carry R-R charges, so they are
sources of (or interact with) R-R fields, and similarly NS-branes
carry NS-NS charges, so they are sources of (or interact
with) NS-NS fields. These fields turn into one another by
S-duality transformation, but this does not implies that the
D-brane can have NS-NS charge, or NS-brane can have R-R charge
because the branes as well as the fields also change into one
another under the S-duality transformation. However, ($p+3$)d string theory admits an exceptional
solution. Namely one can show that ($p+3$)d string theory admits
a solution describing D$p$-brane which, however, contains NS-NS component. To be precise, it is NS-NS component in the sense that it couples with dilaton with a factor $e^{-2\Phi}$
(we will call it simply NS-brane throughout this paper), but the brane is D$p$-brane because it carries an R-R charge.

In this report we will first show that ($p+3$)d string theory really admits such an exceptional solution, and using this solution we will argue that in ($p+3$)d string theory the existence of the NS-brane with negative tension is essential to obtain background geometry $R_2$ or $R_2 / Z_n$ on the transverse dimensions, and the usual codimension-2 brane solutions with these background geometries already contain the negative tension NS-brane implicity in their ansatz. Such an argument leads us, in the context of brane world scenarios, to a conjecture that true background brane immanent in our spacetime may perhaps be NS-brane, rather than D-brane.

We begin with a ($p+3$)d action
\begin{equation}
I_{p+3}= \frac{1}{2\kappa^2} \int d^{p+3}X \sqrt{-G} \, \Big[
e^{-2\Phi} \big[ R + 4 (\nabla \Phi)^2 \big] - \frac{1}{2 \cdot
(p+2)!} F^{2}_{p+2} \Big] \,\,,
\end{equation}
and a brane action
\begin{equation}
I_{brane}= - \int d^{p+1}X \sqrt{-det|g_{\mu\nu}|}\, T_p (\Phi) +
\mu_p \int A_{p+1} \,\,,
\end{equation}
where $\Phi$ is the ($p+3$)d dilaton, and the R-R field strength
$F_{p+2}$ is given by $F_{p+2}=dA_{p+1}$. Also, $g_{\mu\nu}$ is a
pullback of $G_{AB}$ to the $(p+1)$d brane world, thus the first
term of (2) represents the $\sigma$-model term. $T_p (\Phi)$, on the
other hand, represents the tension of the $p$-brane which takes
(at the tree level) the form
\begin{equation}
T_p (\Phi) =T^{(D)}_0 e^{-\Phi}
\end{equation}
if the brane is a D$p$-brane, while it takes
\begin{equation}
T_p (\Phi) =T^{(NS)}_0 e^{-2\Phi}
\end{equation}
if the brane is an NS-brane\footnote{In 10d string theories the tension of the S-dual of a D$p$-brane is generally given by $T_p (\Phi) = T_0 e^{-n \Phi}$ with $n={(p-1)/2}$. So the NS-brane represented by (4) is not an S-dual of the D$p$-brane under discussion unless $p=5$ (see also footnotes 2 and 3). Nevertheless, the existence of such an NS-brane is indispensable for all $p$ to obtain background geometry $R_2$ or $R_2 /Z_n$ on the transverse dimensions as will be discussed later.}, where the constants $T^{(D)}_0$ and
$T^{(NS)}_0$ are both of the order $\sim 1/{\alpha^{\prime}}^{(p+1)/2}$.
The second term of (2) implies that the $p$-brane is electrically coupled with R-R ($p+1$)-form,
and it carries an electric R-R charge $\mu_p$. Also the action (1) only includes the R-R ($p+2$)-form, it does not include NS-NS $n$-form. So (1) typically describes a D-brane in the
usual theories.

Let us introduce a ($p+3$)d metric of the form
\begin{equation}
ds^2_{p+3} = e^{A(\hat{r})} ds^2_3 + e^{B(\hat{r})} d\vec{x}^2_p
\,\,,
\end{equation}
where $d\vec{x}^2_p \equiv dx^2_1 + \cdots + dx^2_p$, and $ds^2_3$
is given by
\begin{equation}
ds^2_3 = -N^2 (\hat{r}) dt^2 + \frac{d\hat{r}^2}{f^2 (\hat{r})}
+R^2 (\hat{r}) d \theta^2 \equiv \hat{g}_{ab} dy^a dy^b \,\,.
\end{equation}
In the above metric $e^{A(\hat{r})}$ is an extra degree of freedom
which could have been absorbed into $ds^2_3$, so it can be taken
arbitrarily as we wish. Also the metric (6) includes one more extra
degree of freedom associated with the coordinate transformation $\hat{r}
\rightarrow \hat{r}^{\prime}=F(\hat{r})$. The ansatz for the $(p+1)$-form
field is given by
\begin{equation}
A_{p+1} (\hat{r}) = \xi (\hat{r}) dt \wedge dx^1 \wedge \cdots
\wedge dx^p ~\rightarrow~ F_{t\hat{r}1 \cdots p} =
\partial_{\hat{r}} \xi \,\,.
\end{equation}
Since the fields do not depend on the coordinates along the
$p$-brane, the above actions can be reduced to the following 3d
effective actions :
\begin{equation}
I_3 = \frac{1}{2\kappa^2} \int  d^3 y \sqrt{- \hat{g}_3}\,\,
\Big[\,\,\hat{R}_3 \,-\,4(\partial \Phi)^2 \,+\,2p(\partial
\Phi)(\partial B)\,- \, \frac{p(p+1)}{4} (\partial B)^2 \, -\,
\frac{1}{2}\,e^{-2\Phi} \hat{g}^{tt} (\partial \xi)^2 \,\Big]\,\,,
\end{equation}
and
\begin{equation}
I_{brane} = -\int  d^3 y \sqrt{- \hat{g}_3}\,\,e^{2\Phi} T_p
(\Phi) \delta^2 (\vec{\hat{r}}) + \mu_p \int  d^3 y \sqrt{
\hat{g}_2}\,\, \xi (\hat{r}) \delta^2 (\vec{\hat{r}}) \,\,,
\end{equation}
where $\hat{R}_3$ is the 3d Ricci-scalar obtained from
$\hat{g}_{ab}$, and the 2d delta-function $\delta^2
(\vec{\hat{r}})$ has been normalized by $\int d^2 \vec{\hat{r}}
\sqrt{\hat{g}_2} \,\delta^2 (\vec{\hat{r}})=1$, where
$\vec{\hat{r}} \equiv (\hat{r}, \theta)$ and $\sqrt{\hat{g}_2}=
\sqrt{-\hat{g}_3}/\sqrt{-\hat{g}_{tt}}$. Also in obtaining the
above 3d action we have chosen
\begin{equation}
A=4\Phi -pB \,\,,
\end{equation}
so that 3d effective dilaton $\phi (=\Phi - (A/4) - p(B/4)$)
vanishes, and the kinetic term for $\hat{g}_{ab}$ becomes the
standard Hilbert-Einstein action.

Let us consider the field equations. It is convenient to consider
the equation for $\xi(\hat{r})$ first. It is given by
\begin{equation}
\frac{1}{\sqrt{\hat{g}_2}} \partial_a \Big[\sqrt{\hat{g}_2}
\frac{\hat{g}^{ab}}{\sqrt{-\hat{g}_{tt}}} e^{-2\Phi} \partial_b
\xi \Big] =2\kappa^2 \mu_p \delta^2 (\vec{\hat{r}}) \,\,,
\end{equation}
which, upon integration, gives
\begin{equation}
\partial_{\hat r} \xi = \frac{\kappa^2}{\pi}\mu_p \, e^{2\Phi}
\sqrt{\frac{-\hat{g}_{tt}
\hat{g}_{\hat{r}\hat{r}}}{\hat{g}_{\theta\theta}}} \,\,,
\end{equation}
and consequently one finds (see (7))
\begin{equation}
\frac{1}{2\kappa^2} \int {}^{*}F_{p+2} = (-)^{p+1} \mu_p \,\,,
\end{equation}
which shows that the $p$-brane located at $\vec{\hat r}=0$ carries
an R-R charge $\mu_p$. The remaining field equations are
\begin{equation}
 N(fR^{\prime})^{\prime} +  NfR \mathcal{H}
 + \frac{1}{4} \frac{fR}{N} e^{-2\Phi} {\xi^{\prime}}^2 \,=\,-\kappa^2 \frac{NR}{f} e^{2\Phi} \,
T_{p}(\Phi)\, \delta^2(\vec{{\hat{r}}}) \,\,,
\end{equation}
\begin{equation}
 N^{\prime}fR^{\prime} - NfR \mathcal{H}  + \frac{1}{4} \frac{fR}{N} e^{-2\Phi} {\xi^{\prime}}^2\,=\, 0,
\end{equation}
\begin{equation}
(N^{\prime}f)^{\prime}R + NfR \mathcal{H} - \frac{1}{4}
\frac{fR}{N} e^{-2\Phi} {\xi^{\prime}}^2\,=\, 0 \,\,,
\end{equation}
\begin{equation}
(NfR\Phi^{\prime})^{\prime} - \frac{(p+1)}{8} \frac{fR}{N}
e^{-2\Phi} {\xi^{\prime}}^2 \,=\, \frac{(p+1)}{2}\kappa^2
\frac{NR}{f} e^{2\Phi} \,\Big[ T_{p}(\Phi) +
\frac{1}{2}\frac{\partial T_{p}(\Phi)}{\partial \Phi} \Big]
\,\delta^2 (\vec{{\hat{r}}}) \,\,,
\end{equation}
\begin{equation}
(NfRB^{\prime})^{\prime} -\frac{1}{2} \frac{fR}{N} e^{-2\Phi}
{\xi^{\prime}}^2\,=\, 2\kappa^2 \frac{NR}{f} e^{2\Phi} \Big[
T_{p}(\Phi) + \frac{1}{2}\frac{\partial T_{p}(\Phi)}{\partial
\Phi}\Big] \,\delta^2 (\vec{{\hat{r}}})\,\,,
\end{equation}
where $\mathcal{H} \equiv 2 {\Phi^{\prime}}^2 -p\Phi^{\prime}
B^{\prime} + \frac{p(p+1)}{8} {B^{\prime}}^2$, and the "prime"
denotes the derivative with respect to $\hat{r}$.

Among the above equations, the first three follow from the 3d
Einstein equations, while the last two are linear combinations
of the equations for $\Phi$ and $B$. These five equations
constitute a complete set of linearly independent equations of
motion. However, a linear combination of (15) and (16) gives
\begin{equation}
 (fR{N^\prime})^\prime \,=\,0\,\,,
\end{equation}
while (17) and (18) imply
\begin{equation}
B = \frac{4}{(p+1)}\Phi \,\,.
\end{equation}
So due to (19) and (20) the number of independent equations
reduces only to three, and from (10) and (20) the $(p+3)$d metric
becomes
\begin{equation}
ds{_{p+3}^2}  = e^{4\Phi/(p+1)} \Big[\big(\frac{d\hat{r}^2}{f^2} +
R^2 d\theta^2\big)+\big( - N^2 dt^2 + d \vec{x}{_p^2} \,\big)\Big]
\,\,.
\end{equation}
In (21), the usual black $p$-brane solution may be obtained by taking $N(\hat{r})=f(\hat{r})$.
For $N=f$, (19) is written as
\begin{equation}
(f^2 )^{\prime} = \frac{b_1}{R}\,\,,~~~~~(b_1 = {\rm const.})\,\,,
\end{equation}
and if we introduce a new coordinate $r$ defined by $d\hat{r}/R =dr/r$, (22) is
immediately solved by
\begin{equation}
f^2 =b_0 + b_1 \ln r \,\,,~~~~~(b_0 = {\rm const.})\,\,.
\end{equation}
In the present paper we are interested in the extremal type solution which
preserves the ($p+1$)d Poincar\'{e} invariance. So we take $b_0 =1$, $b_1 =0$;
i.e., $f=N=1$. In most cases such an extremal solution possesses maximal unbroken supersymmetry and
corresponds to a BPS state.

In terms of the variable $r$, the metric takes the form
\begin{equation}
ds{_{p+3}^2}  = e^{4\Phi/(p+1)} \Big[\frac{R^2}{r^2}\big({d{r}^2} +
r^2 d\theta^2\big)+\big( - dt^2 + d \vec{x}{_p^2} \,\big)\Big]
\,\,,
\end{equation}
and omitting (16) and (18) one finds that the set of three linearly
independent equations can be written as
\begin{equation}
\nabla^2 \ln R + \frac{q^2}{2}\,\psi = -\frac{1}{2} C_1 \,
\delta^2 (\vec{r})\,\,,
\end{equation}
\begin{equation}
\nabla^2 \Phi -\frac{(p+1)}{8} \,q^2 \psi = \frac{(p+1)}{4} C_2
\,\delta^2 (\vec{r})\,\,,
\end{equation}
\begin{equation}
~~~\big( \frac{d\Phi}{dr}\big)^2 = \frac{(p+1)}{8} \,q^2 \psi
\,\,,
\end{equation}
where $q \equiv \kappa^2 \mu_p/\pi$, and $\nabla^2$ is the flat
space Laplacian $\nabla^2 \equiv (1/r)(d/dr)(rd/dr)$ (so $\delta^2
(\vec{r})$ is now normalized by $\int r dr d\theta \delta^2
(\vec{r})=1$). Also $\psi$ and $C_i$ are defined, respectively, by
\begin{equation}
\psi = \frac{e^{2\Phi}}{r^2} \,\,,
\end{equation}
and
\begin{equation}
C_1 = 2\kappa^2 e^{2\Phi} T_{p}(\Phi)\Big|_{\vec{r}=0}\,\,,~~~~~
C_2 = 2\kappa^2 e^{2\Phi} \Big[ T_{p}(\Phi) +
\frac{1}{2}\frac{\partial T_{p}(\Phi)}{\partial \Phi}
\Big]\Big|_{\vec{r}=0} \,\,.
\end{equation}

From (26) and (28) one finds that $\psi$ must satisfy
\begin{equation}
\nabla^2 \ln\psi -\frac{(p+1)}{4} \,q^2 \psi = 4\pi (\alpha -1)
\,\delta^2 (\vec{r})\,\,,
\end{equation}
with
\begin{equation}
\alpha = \frac{(p+1)}{8 \pi} C_2 \,\,,
\end{equation}
and using (30) one can show that the solution to (25) and (26)
 (also see (20)) takes the form
\begin{equation}
R=i_R \, (\psi r^2 )^{k_R} R_0 \,\,, ~~~~e^{\Phi} =(\psi r^2 )
^{k_\Phi}\,\,, ~~~e^B = (\psi r^2 )^{k_B} \,\,,~~~~(R_0 = {\rm
const.})\,\,,
\end{equation}
where $k_M$ ($M \equiv R, \Phi, B$) are
\begin{equation}
k_R = -\frac{2}{(p+1)}\,\,,~~~k_\Phi = \frac{1}{2} \,\,, ~~~ k_B =
\frac{2}{(p+1)} \,\,,
\end{equation}
while $i_R$ is defined by
\begin{equation}
\nabla^2 \ln i_R = 2\pi \beta \, \delta^2 (\vec{r}) \,\,,
\end{equation}
where
\begin{equation}
\beta = -\frac{1}{4\pi} C_1 + \frac{1}{2\pi} C_2 \,\,.
\end{equation}
The solution to (30) is \cite{8}
\begin{equation}
\psi (r)= \frac{a_0}{r^2 \big[ ({r}/{r_0})^{\alpha} -
({r}/{r_0})^{-\alpha} \big]^2}\,\,, ~~~~~~\big(a_0 \equiv
\frac{32}{(p+1)}\frac{\alpha^2}{q^2}\,,~~~ r_0 = {\rm const.}
\big) \,\,,
\end{equation}
while from (34)
\begin{equation}
i_R (r) = \Big( \frac{r}{\tilde{r}_0}\Big)^{\beta} \,\,.
\end{equation}
So the metric (24) now becomes
\begin{equation}
ds^2_{p+3} = ( r/\tilde{r}_0 )^{2(\beta -1)} (\psi
r^2)^{-2/(p+1)} \big( dr^2 + r^2 d\theta^2 \big) + (\psi
r^2)^{2/(p+1)} \big( -dt^2 + d\vec{x}^2_p \big) \,\,,
\end{equation}
where without loss of generality we have identified the constant
$R_0$ in (32) with the constant $\tilde{r}_0$ in (37). The metric (38) is perfectly
well-defined for $\beta \geq 0$. Except the logarithmic singularity arising from the
conformal factor $(\psi r^{2})^{-2/(p+1)}$ (see (40)), it has only a conical singularity at $r=0$
for $\beta \neq 1$ (see (43)).

Though the solution (32) (together with (36) and (37)) satisfies
(25) and (26), we still need for consistency to check whether it
satisfies (27) either. Substituting (32) and (36) into (27) gives
a condition
\begin{equation}
\alpha=0 ~~\rightarrow~~ e^{2\Phi} \Big[ T_{p}(\Phi) +
\frac{1}{2}\frac{\partial T_{p}(\Phi)}{\partial \Phi}
\Big]\Big|_{\vec{r}=0} =0 \,\,,
\end{equation}
and due to this condition (36) reduces to
\begin{equation}
\psi (r)= \frac{\hat{a}_0}{r^2 \big[ \ln({r}/{r_0}) \big]^2}\,\,,
~~~~~~\big(\hat{a}_0 \equiv \frac{8}{(p+1)}\frac{1}{q^2} \,) \,\,,
\end{equation}
(one can also check, by directly substituting (40) into (30), that (40) is really the solution to
(30) for $\alpha =0$), or if we set $r_0 = \exp[\sqrt{8/(p+1)}\,\,(c_0 /q)]$ it can be rewritten as
\begin{equation}
\psi r^2 = \Big[ c_0 - \sqrt{\frac{(p+1)}{8}}\,\frac{\kappa^2 \mu_p}{\pi}
\ln r \Big]^{-2} \,\,,~~~~~~(c_0={\rm const.})\,\,.
\end{equation}
(41) is a typical form of the 2d Green's function and one finds that
in a particular case (i.e., for $\beta =1$ and for the total dimensions
$p+3=10$) (38) precisely reduces (upon taking $\mu_p =T_0$ in (41)) to the well known codimension-2 brane
solution in \cite{9} (see (3.36) of ref.\cite{9}).

The tension (3) satisfies (39) because $\alpha$ becomes
$\sim \kappa^2 e^{\Phi}\, T_0^{(D)}$ for (3), and $e^{\Phi}$ goes to zero
as $\vec{r} \rightarrow 0$ as can be seen from (28) and (40).
In fact, $C_1$ and $C_2$ vanish for (3) because they are both proportional
to $e^{\Phi}$, and consequently $\alpha$ and $\beta$ also vanish for (3).
Since $C_1$ and $C_2$ vanish, the solution to the equations (25)-(27)
becomes nonsingular and "solitonic". However, the 2d transverse space($\equiv \Sigma_2$)
defined by ($r$, $\theta$) becomes a cylinder $R_1 \times S_1$ for $\beta =0$ as can be observed from (38). The radius of the cylinder is $\tilde{r}_0$ which, however, must be taken to be zero because only for this value of $\tilde{r}_0$ the transverse space $\Sigma_2$ admits a codimension-2 brane at $r=0$. This is in contrast to the situation of the D7-brane in F-theory, where the transverse space becomes cylindrical when the number of D7-brane is $12$. But in that case the tip of the cylinder is not sharp-pointed, and therefore the radius of the cylinder takes a nonzero value there. So if we want to have a nonzero $\tilde{r}_0$, we need to blunt the tip of the cylinder just as in the D7-brane solution of the F-theory. Turning back to the metric (38) the cylinder spreads out to become $R_2$ or $R_2 /Z_n$ if we introduce a negative tension NS-brane at $r=0$ as we shall see in the followings.

Let us turn to the NS-brane. We observe that the tension (4) of the NS-brane also satisfies (39). It strictly satisfies (39) for arbitrary
$T_0^{(NS)}$ and $e^{\Phi}$ due to its particular functional dependence on $\Phi$. But the coefficient
$C_1$ and consequently $\beta$ do not vanish this time. They are now independent of $e^{\Phi}$, and only determined by $T_0^{(NS)}$; namely
\begin{equation}
\beta = - \frac{\kappa^2}{2 \pi} T^{(NS)}_0 \,\,.
\end{equation}
The effect of the nonvanishing $\beta$ manifests itself in the metric (38).
In (38), $\beta$ is related with a deficit angle of $\Sigma_2$. Introducing a new variable $\rho$ defined by $\rho/\rho_0 \equiv (r/\tilde{r}_0)^{\beta}$ (and choosing $\rho_0 = \tilde{r}_0 / \beta$) one finds
\begin{equation}
ds^3_{p+3} = (\psi r^2)^{-2/(p+1)} (d \rho^2 + \beta^2 \rho^2 d \theta^2 ) + (\psi r^2)^{2/(p+1)} (-dt^2 + d \vec{x}_p^2 ) \,\,,
\end{equation}
so the deficit angle $\delta$ is given by $\delta = 2\pi (1- \beta )$. If $\beta=1$, $\Sigma_2$ is simply (locally) $R_2$. But if $\beta =1/n$, $\Sigma_2$ becomes an orbifold $R_2 /Z_n$ with an orbifold singularity at $\vec{\rho}=0$. Also $\beta$ must be positive in order for $\delta$ not to exceed $2 \pi$ (this condition coincides with that imposed on the metric (38)). So in this case the NS-brane should be a negative-tension brane (i.e., $T_0^{(NS)} <0$) as one can see from (42). If $\beta <0$, however, $T^{(NS)}_0$ is positive. But in this case $\delta$ exceeds $2\pi$, and $\Sigma_2$ turns into compact space. We will briefly consider this case later.

The existence of the NS-brane with negative tension is essential to obtain the background geometry $R_2$ or $R_2 /Z_n$ on the space $\Sigma_2$. Recall that the geometry of $\Sigma_2$ was a cylinder $R_1 \times S_1$ for $\beta=0$. It spreads out to become $R_2$ or $R_2 /Z_n$ when $\beta$ takes a nonzero positive value, or equivalently when the negative tension NS-brane is introduced at $\vec{\rho}=0$. In the absence of the D$p$-brane ($\mu_p =0$), (43) represents the vacuum with a flat geometry $R_2$ when $T^{(NS)}_0 = -2\pi /\kappa^2$ ($\beta =1$), while it represents an orbifold $R_2 /Z_n$ when $T^{(NS)}_0 = -2\pi /n\kappa^2$ ($\beta =1/n$). So formally, $R_2$ or $R_2 /Z_n$ is equivalent to a cylinder $R_1 \times S_1$ plus a negative tension NS-brane placed at $\vec{\rho}=0$, and it is conjectured that the usual $(p+3)$d solutions with background geometry $R_2$ or $R_2 /Z_n$ already contain the negative tension NS-brane implicity in their ansatz.

The above argument is supported by the fact that the metric (6) (together with (5)) is the most general ansatz we can think of for the 3d subsector of the $(p+3)$d metric. Note that it admits an extra equation (i.e., the equation for $R(\hat{r})$), which does not exist in the case of the usual $(p+3)$d ansatz with a fixed geometry $R_2$ or $R_2 /Z_n$ on the transverse dimensions. This suggests that the solution obtained from the ansatz (6) would be the one that is closer to the true extremum of the action than the others. Note that fixing an ansatz generally leads to a limited class of geometries. According to our discussion fixing the geometry of the ansatz corresponds to fixing the value of $\beta$ from the beginning. So for instance the ansatz with a geometry $R_2$ on $\Sigma_2$ corresponds to an ansatz with a fixed value $\beta=1$ or $C_1 = -4 \pi$ (recall that $C_2$ vanishes for both D-brane and NS-brane), and with this value of $C_1$  (25) turns into a "solitonic" equation
\begin{equation}
\bigtriangledown ^2 \ln \hat R + \frac{q^2}{2} \psi =0 ~~~~~~~~{\rm with} ~~
\hat{R} \equiv R/r \,\,.
\end{equation}
Observe that the delta function term, being absorbed into $\hat{R}$, does not appear in (44).

It is not clear what makes $T^{(NS)}_0$ to take those particular values, i.e., $T^{(NS)}_0 = - 2\pi / \kappa^2$ for $R_2$ and $T^{(NS)}_0 = -2\pi /n\kappa^2$ for $R_2 /Z_n$. One of the answer to this question may be found from the supersymmetry preserved by the spacetime.  Obviously, the vacuum with a flat geometry $P_{p+1} \times R_2$, where $P_n$ represents $n$-dimensional Poincar$\acute{\rm e}$ space, preserves full supersymmetry of the theory whose supercharges are given by the spinorial representation of $SO(p,1) \times SO(2)$, and it is also known that orbifolds generally preserve (a part of) supersymmetry under certain conditions \cite{4}. Since these spaces are stable by the supersymmetry, it may be conjectured that $T^{(NS)}_0$ prefers those particular values above all others.

It is known that negative energy objects generally lead to instabilities or unusual gravitational effects \cite{10}. However, some negative tension objects such as orientifolds \cite{11} of the string theory are very well-defined \cite{12} and even find good applications \cite{13}. Indeed, negative tension brane fixed at orbifold fixed point is generally known to be free from such instability problems, which is a similar configuration as our case where a negative tension NS-brane is fixed at the orbifold fixed point $\vec{\rho}=0$. So the action does not include the pathological negative-definite kinetic energy term which causes an instability of the negative tension objects\cite{13-1}. The literatures supporting the stability of the negative tension branes can be found in \cite{14}, and in particular the stability of the codimension-2 negative tension brane has been discussed in \cite{15}.

Apart from this, our negative tension brane has a special feature. As mentioned previously the role of the negative tension NS-brane is just to spread out $R_1 \times S_1$ to convert it into $R_2$ or $R_2 /Z_n$. Once $\Sigma_2$ becomes $R_2$ or $R_2 /Z_n $, the negative tension NS-brane essentially disappears in compensation for it. It is absorbed into the background space and does not show up anymore (see (44)). So $\Sigma_2$ simply appears as $R_2$ whose ADM mass is just zero, or the orbifold $R_2 /Z_n$ whose ADM mass is positive, implying that there is no negative energy object in the space. The negative energy object only appears for $\beta >1$. For $\beta >1$, the deficit angle and the ADM mass of $\Sigma_2$ are both negative. Since no negative energy object appears in the space (namely since they have been absorbed into $R_2$ or $R_2 /Z_n $), the stability problem of the negative tension NS-brane reduces to the stability problem of the background space $R_2$ or $R_2 /Z_n$ which, however, known to be both supersymmetric and stable.

Though the negative tension NS-brane does not manifest itself in the space $R_2$ or $R_2 /Z_n$, its effect on the geometry of these spaces is crucial. In order to see this easily it is convenient to assume that the D-brane and NS-brane are seperately well-defined. Now consider a configuration that a D-brane is introduced at the orbifold singularity of $R_2 /Z_n$, or at the orgin of $R_2$. This amounts to increasing $C_1$ by $\delta C_1$, and consequently $\beta$ by $\delta \beta$. However, since NS-brane is much heavier than D-brane the increment $\delta C_1$ (or $\delta \beta$) will be very small as compared with $C_1$, and the ratio $\delta C_1 /C_1$ (or $\delta \beta/\beta$) is in fact of an order $\sim e^{\Phi}$ which goes to zero as $e^{\Phi} \rightarrow 0$. If we consider, on the other hand, the change $\delta C_1 (\delta \beta)$ due to quantum fluctuations of SM fields with support on the D-brane, the ratio $\delta C_1 /C_1$ gets even smaller; it is of an order $\sim e^{2\Phi}$ \cite{5}, which implies that the geometry of $\Sigma_2$ is virtually unaffected by the quantum fluctuations of SM fields living on the D-brane. In the brane world models the intrinsic curvature of the brane is {\it a priori} zero. So the whole quantum fluctuations of SM fields entirely contribute to changing the bulk geometry, and hence the bulk geometry is generally disturbed severely by the quantum fluctuations. In our case, however, the disturbance due to quantum fluctuations is highly suppressed as mentioned above. The bulk geometry, as well as the flat geometry of the brane, is practically insensitive to the quantum fluctuations and such a feature is essential to addressing the cosmological constant problem. Namely it provides a new type of self-tuning mechanism with which to solve the cosmological constant problem \cite{5}.

The solution obtained in this paper has an unusual property. It contains both NS-brane and D-brane components ; it
couples to the gravity with a factor $e^{-2\Phi}$, while it carries an R-R charge. The NS-brane component is essentially used to obtain the background geometries like $R_2$ or $R_2 /Z_n$. So once these background geometries are obtained from $R_1 \times S_1$, the NS-brane hides itself behind the background space and we are only left with a D-brane with an R-R charge. This may not be distinguished from the usual configurations where a D-brane is placed on the background space $R_2$ or $R_2 /Z_n$ from the beginning.
To find the relation of our solution to the existing solutions let us dualize $F_{p+2}$ to the magnetic 1-form $F_1 (\equiv {}^\ast F_{p+2})$ in (1). It was shown in \cite{16} that the action (1) plus (2) is classically equivalent to the action
\begin{equation}
I_{p+3} = \frac{1}{2\kappa^2} \int d^{p+3}X \sqrt{-G}\, \big[\, e^{-2\Phi} [R + 4 (\nabla \Phi)^2 ] - \frac{1}{2} F^2_1 \big]
\end{equation}
plus
\begin{equation}
I_{brane} = -\int d^{p+1}X \sqrt{-det|g_{\mu\nu}|} \, T_p (\Phi) \,\,.
\end{equation}
Thus field equations are basically the same as before except that (11) is replaced by the equation for $F_1$ :
\begin{equation}
\partial_{\theta} \big( \sqrt{- \hat{g}_3}\, e^{2\Phi} \hat{g}^{\theta\theta} F_{\theta} \, \big) \,=\,0 \,\,.
\end{equation}
The solution satisfying both (47) and the Bianchi identity $\partial_{[a}F_{b]}=0$ is $F_{\theta} = {\rm constant} \equiv {\kappa^2 \mu_p}/{\pi}$. So $F_1$ satisfies
\begin{equation}
\frac{1}{2\kappa^2} \int_{S_1} F_1 \,=\,(-)^{p+1} \mu_p \,\,,
\end{equation}
where $S_1$ is a circle around the origin where the $p$-brane is located. Observe that (48) is consistent with (13), but $\mu_p$ is now a magnetic charge because the brane is a magnetic source under $F_1$. Since $F_1 =da$, $a$ representing the axion field, we get
\begin{equation}
a=(-)^{p+1} \, \frac{\mu_p}{2\pi} \,\theta \,= \, (-)^{p+1} \, q \, \theta\,\,,
\end{equation}
where we have set $2\kappa^2 =1$. Since field equations are essentially the same as before, so also is the solution.  In the Einstein frame it is given (see (38) and (41)) by
\begin{equation}
ds^2_{p+3} = -dt^2 + d \vec{x}_p^2 + (\frac{r}{\tilde{r}_0})^{-\delta/\pi}{h(r)}^{8/(p+1)}[dr^2 +  r^2 d \theta^2 ]  \,\,,
\end{equation}
\begin{equation}
e^{\Phi} = {h(r)}^{-1} \,\,,
\end{equation}
where
\begin{equation}
h(r) = \big[ c_0 - \sqrt{\frac{(p+1)}{8}} \frac{\mu_p }{2\pi}\ln r \big] \,\,.
\end{equation}

Let us now return to the case of $p=7$ since in that case we have a full 10d string theory\footnote{From the 10d string theoretical point of view, the ($p+3$)d action of this paper with $p<7$ may be identified with the reduced action in \cite{16-1} (with the additional scalar truncated), which gives a family of deformed D$p$-brane solutions.} and our solution may be relatively well understood in relation to the D7-brane or $(p,q)$ sevenbrane of the type IIB theory. For the 7-branes it is customary to use the complex coordinate defined by $z=r e^{i\theta}$. In the Einstein frame (45) can be rewritten as
\begin{equation}
I_{10}= \int d^{10}X \sqrt{-G} \, \big[\, R - \frac{\partial\tau \bar\partial \bar{\tau}}{2 \tau^2_2 } \,\big] \,\,,
\end{equation}
where $\tau=a + i e^{-\Phi} \equiv \tau_1 + i \tau_2$. The equation for $\tau$ following from this action takes the form
\begin{equation}
\partial \bar\partial \tau + 2 \frac{\partial\tau \bar\partial\tau}{\bar \tau - \tau} =0 \,\,,
\end{equation}
which is trivially solved by any holomorphic or antiholomorphic function of $z$.
Also for such $\tau$, and for an ansatz
\begin{equation}
ds^2 = -dt^2 + d \vec{x}_7^2 + H(z, \bar{z})dz d\bar{z}  \,\,,
\end{equation}
the Einstein equations take the form
\begin{equation}
\partial \bar\partial \ln H = \partial \bar\partial \ln \tau_2 \,\,.
\end{equation}

Though (54) is trivially solved by any holomorphic or antiholomorphic function of $z$, $\tau$ is not allowed to be any arbitrary complex number. Since (53) has an $SL(2,\mathbb{Z})$ invariance (at the quantum level), and the $SL(2,\mathbb{Z})$ modular transformation of $\tau$ reproduces the same torus, $\tau$ is restricted to lie in the fundamental domain. A well-known modular invariant solution to (56) is the D7-brane solution of the F-theory given by \begin{equation}
H(z, \bar{z})\,=\,\tau_2 \eta^2 \bar{\eta}^2 |z^{-\frac{N}{12}}|^2  \,\,,
\end{equation}
where $\eta$ is Dedekind's function and $N$ is the number of D7-branes located at $z=0$. Also a suitable choice for $\tau (z)$ would be
\begin{equation}
\tau (z) \sim \frac{N}{2\pi i} \, \ln z
\end{equation}
near $z=0$. Observe that (58) is consistent with (49) and (51) if we set $\mu_p =N$. With this choice for $\tau$, encircling the point $z=0$ induces the monodromy $\left(\begin{array}{rr} 1&1\\0&1 \end{array}\right) \in SL(2,\mathbb{Z})$ for $N=1$, which ensures that the brane at $z=0$ is a $(1,0)$ sevenbrane, i.e., D7-brane.
This D7-brane solution is distinguished from (50) by the following facts. First, the scale factor $H(z, \bar{z})$ is so adjusted that it never vanishes at $z=0$, and therefore it is regular at $z=0$ while (50) has a conical singularity there. The conical singularity at $z=0$ seems to be unavoidable if one insists that the metric should be rotationally symmetric. It was argued in \cite{17} that in general the magnetic 7-brane solution preserving $P_8 \times SO(2)$ symmetry must be of the form (50). Note that (57) is not rotationally symmetric.

Though these two solutions do not coincide near $z=0$, they asymptotically compatible with one another if we take $\delta$ properly. At infinity (57) becomes $H(z, \bar{z}) \, \sim \, \tau_2 \, (z\bar{z})^{-N/12}$, which accords with (50) if we take $\delta =\frac{\pi}{6}N$. Also since (49) produces the same monodromy as (58) for $\mu_p =1$, our solution described by (50) essentially corresponds to a D7-brane solution. For these reasons we may say that (50) is a rotationally symmetric version of (57), or conversely (57) is a regularized version of (50). This suggests that they are basically of the same kind both describing D7-branes, which then leads us to a conjecture that the background space of the D7-brane described by (55), and consequently the background space of the $(p,q)$ sevenbranes either, the $SL(2,\mathbb{Z})$ family of the D7-brane, all contain the negative tension NS-brane implicity in their ansatz just as in the case of the rotationally symmetric solution (50). This conjecture can be immediately extended to the brane world scenarios in the context of F-theory. Namely in F-theory, the D7-brane or the $(p,q)$ sevenbrane wrapped on a 4-cycle corresponds to the 3d space where we live, and therefore the conjecture is that true background $p$-brane immanent in our spacetime may perhaps be NS-NS type brane, rather than D-brane.

Apart from this, it is interesting to observe that an individual D7-brane of the F-theory also contains an NS-NS component in itself.\footnote{This NS-NS component is different from the one that results from the $SL(2,\mathbb{Z})$ transformation of the D7-brane. Recall that the monodromy generated by the 7-brane under discussion is anyhow $\left(\begin{array}{rr} 1&1\\0&1 \end{array}\right)$, not $\left(\begin{array}{rr} 1-pq & p^2 \,\,\,\, \\-q^2 \,\,\, & 1+pq \end{array}\right)$ of the $(p,q)$ sevenbrane. Also S-dual of the D7-brane (for instance the exotic NS7-brane \cite{18}) has a tension of the form $T \sim T^{(NS)}_0 \, e^{-3\Phi}$ instead of (4). See the next lines.} Recall that the D-brane tension (3) can not generate a deficit angle of $\Sigma_2$. But we see that (57) exhibits a deficit angle $\delta = {\pi}/6$ at infinity, which suggests that the D7-brane described by (57) may carry an NS-brane component with a tension given by (4). But in this case $T^{(NS)}_0$ associated with an individual D7-brane is positive on the contrary to the case of the background space. Such a feature of the D7-brane is entirely due to the factor $|z^{{-N}/12}|^2$ in (57), which has been introduced to avoid the singularity at $z=0$ \cite{19}, but gives a deficit angle at infinity instead. Since the deficit angle due to $N$ D7-branes is given by $\delta =\frac{\pi}{6}N$ (or $\beta = 1-\frac{N}{12}$), if we apply (42) to these D7-branes the corresponding $T_0^{(NS)}$ becomes
\begin{equation}
T^{(NS)}_0 \,=\, - \frac{4\pi}{2 \kappa^2}\, (1-\frac{N}{12}) \,\,.
\end{equation}
So if $N<12$, $T^{(NS)}_0$ is negative, and this corresponds to the case where the deficit angle is less than $2\pi$ and the 2d transverse space $\Sigma_2$ is non-compact. Also if $N=12$, $T^{(NS)}_0$ vanishes, and in this case the deficit angle is precisely $2\pi$, so $\Sigma_2$ is a cylinder $R_1 \times S_1$. Finally if $N>12$, $T^{(NS)}_0$ becomes positive, and this corresponds to the case where $\delta$ exceeds $2\pi$. In this case the transverse space $\Sigma_2$ is "eaten up" by the D7-branes and it becomes compact. In particular, if $N=24$, which corresponds to $\delta = 4\pi$, $\Sigma_2$ becomes $S_2$.\footnote{The radius of $S_2$ would be equal to $\tilde{r}_0$ which, however, is zero in our case unless the conical singularity at $\vec{r}=0$ has been blunted.}

In (59), $N=0$ gives $T_0^{(NS)}= -{4\pi}/{2\kappa^2}$, which corresponds to $\beta=1$, and therefore the background space $R_2$ without any D7-brane.  In general case, in the absence of the $p$-branes $\beta$ must be of an order $\beta \sim 1$ in order for $\Sigma_2$ to have a (local) background geometry $R_2$ or $R_2/Z_n$. Since $\kappa^2 \sim 1/M_{p+3}^{p+1}$ and $|T_0^{(NS)}| \sim M_s^{p+1}$ (where $M_{p+3}$ is the ($p+3$)d  Planck scale), $\beta \sim 1$ implies $M_s /M_{p+3} \sim 1$. This accords with the hierarchy assumption that there exists only one fundamental short distance scale (i.e., the electroweak scale $m_{EW}$) in nature, which again supports the conjecture that the a natural background $p$-brane of our universe would be NS-NS type brane, rather than D-brane.

\vskip 1cm
\begin{center}
{\large \bf Acknowledgement}
\end{center}

This work was supported by Kyungsung University in 2005.

\end{document}